\documentclass{article}

\usepackage{spconf,amsmath,graphicx,subfigure,hyperref}

\title{Branchy-GNN: a Device-Edge Co-Inference Framework for Efficient Point Cloud Processing}
\name{Jiawei Shao$^{ \dagger}$, Haowei Zhang$^{\star}$, Yuyi Mao$^{\dagger}$, and Jun Zhang$^{ \dagger}$}
\address{$^{\dagger}$The Hong Kong Polytechnic University, Hong Kong\\
$^{\star}$The Hong Kong University of Science and Technology, Hong Kong}

\begin{document}
%
\maketitle
\begin{abstract}


The recent advancements of three-dimensional (3D) data acquisition devices have spurred a new breed of applications that rely on point cloud data processing. However, processing a large volume of point cloud data brings a significant workload on resource-constrained mobile devices, prohibiting from unleashing their full potentials. Built upon the emerging paradigm of device-edge co-inference, where an edge device extracts and transmits the intermediate feature to an edge server for further processing, we propose Branchy-GNN for efficient graph neural network (GNN) based point cloud processing by leveraging edge computing platforms. In order to reduce the on-device computational cost, the Branchy-GNN adds branch networks for early exiting. Besides, it employs learning-based joint source-channel coding (JSCC) for the intermediate feature compression to reduce the communication overhead. Our experimental results demonstrate that the proposed Branchy-GNN secures a significant latency reduction compared with several benchmark methods.

\end{abstract}
\section{Introduction}
\label{introduction}

Point cloud is one of the most important data formats for three-dimensional (3D) objects and space representations. With the recent development of fast 3D data acquisition devices, such as Lidar scanners and stereo cameras, there has been a significant uprise in mobile applications for graphics and vision that directly process raw point clouds.
Although some hand-crafted features on point clouds have been proposed for decades, the recent success of deep neural networks (DNNs) \cite{krizhevsky2012imagenet} inspires us to learn the point cloud features by neural networks for better performance.
Nonetheless, applying deep learning for point cloud feature extraction is highly non-trivial due to its irregular grid.
On the one hand, some prior studies \cite{maturana2015voxnet,riegler2017octnet} convert the point cloud into 3D grids for 3D convolution. This stream of methods tends to cause information loss in quantization.
On the other hand, the pointwise based methods \cite{qi2017pointnet,qi2017pointnet++} allow an unordered set of points as input without mapping them to a grid.
These methods, however, adopt a shared multilayer perceptron (MLP) to process the points independently at the local subset and neglect the geometric information.

Graph neural networks (GNNs) \cite{wang2019dynamic_DGCNN}, which capture the dependence of graphs via message passing between neighboring nodes, are powerful for point cloud processing.
The point cloud data can adopt a graph as its compact representation after generating edge features.
For effective processing, GNNs exploit the relationships between a point and its neighbors in the Euclidean space and extract fine-grained features.
GNN-based point cloud processing has empowered many new applications, including indoor navigation \cite{diaz2016indoor}, self-driving vehicles \cite{yue2018lidar_driving}, and shape modeling \cite{chen2019learning_shape_modeling}.
The wide deployment of smart edge devices with 3D data acquisition capability necessitates GNN-based inference at the wireless network edge.
In practice, \emph{Device-only inference}, where the model computations are executed locally, and \emph{edge-only inference}, where the data is transmitted to the edge server for remote processing, are two conventional inference strategies.
Nevertheless, neither of them can cater to a stringent latency requirement: The device-only inference suffers from long on-device computation latency due to the computation-intensive models. In contrast, the edge-only inference incurs excessive communication overhead caused by the large volume of input data.
Fortunately, the \emph{device-edge co-inference} paradigm \cite{shao2020communication,shi2020communication_jun}, which forwards an intermediate feature to the edge server for processing, is a promising candidate to reduce the inference latency by striking a balance between the computation and communication overhead.

Although device-edge co-inference effectively reduces the inference latency, most of the existing works focused on DNN-based edge inference \cite{li2018edge_Edgenet,Shi2019ImprovingDC_2step}, which cannot be easily extended to GNN-based inference because of the more severe \emph{data amplification} \cite{kang2017neurosurgeon}, i.e., the intermediate feature size is much larger than the input data.
As a result, dedicated designs for GNN-based inference are needed, which motivates our investigation in this paper.
In particular, we will propose a low-latency co-inference framework, namely, the Branchy-GNN, for effective point cloud processing, which implements an early exit mechanism to save on-device computation. In addition, the learning-based joint source-channel coding (JSCC) \cite{bourtsoulatze2019deepJSCC_deniz} is applied to the intermediate features in the branch networks for communication latency reduction.
Our experimental results validate the effectiveness of the proposed framework in reducing the edge inference latency for GNNs under different wireless communication environments.

The rest of this paper is organized as follows. In Section \ref{Prior Work}, we introduce the preliminaries on GNN-based point cloud processing and identify the unique challenges. In Section \ref{proposed}, we present the proposed Branchy-GNN framework for low-latency inference. Experimental results are presented in Section \ref{experiment}, and we conclude this paper in Section \ref{conclusion}.

\section{Preliminaries}
\label{Prior Work}

\subsection{GNN-based Point Cloud Processing}

GNNs have attracted increasing attention in recent years due to their strength in handling irregular data.
These works \cite{qi2017pointnet,wang2019dynamic_DGCNN} adopt graphs as compact representations of point clouds and leverage GNNs for the high-level feature extracting.
Compared with other point cloud processing techniques like pointwise methods \cite{qi2017pointnet} and volumetric-based methods \cite{maturana2015voxnet}, GNNs excel in exploiting the geometric information and extracting high-resolution features without incurring high memory usage and computational cost.
Particularly, GNNs consider each point as a vertex in a graph and generate edge features based on its neighbors, and the feature learning is performed in the graph domains.
For example, Point-GNN \cite{shi2020point_gnn} encodes the point cloud in a fixed radius near-neighbors graph and updates point features based on the input graph.
In DGCNN \cite{wang2019dynamic_DGCNN}, the graph structure is dynamically updated based on the distances in the semantic space, and the learned features are aggregated based on the new graph.
However, although these methods are effective in point cloud processing, deploying large-scale GNNs to the network edge is non-trivial.

\begin{figure}[h]
\subfigure[The model splitting framework.]{
\begin{minipage}[]{1\linewidth}
\centering
\includegraphics[width=0.95\textwidth]{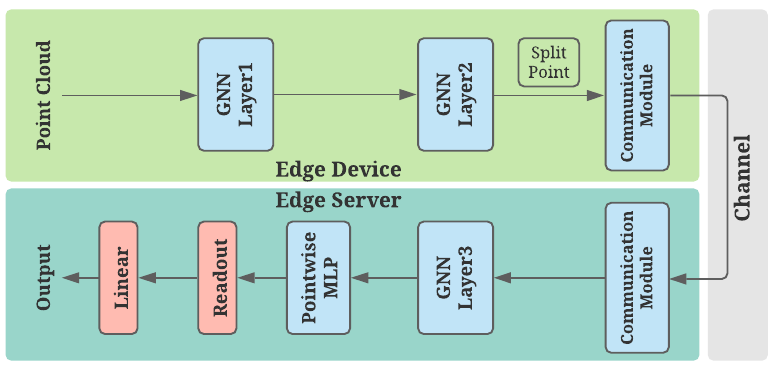}
\label{model splitting}
\end{minipage}%
}%

\subfigure[The Branchy-GNN framework.]{
\begin{minipage}[]{1\linewidth}
\centering
\includegraphics[width=0.95\textwidth]{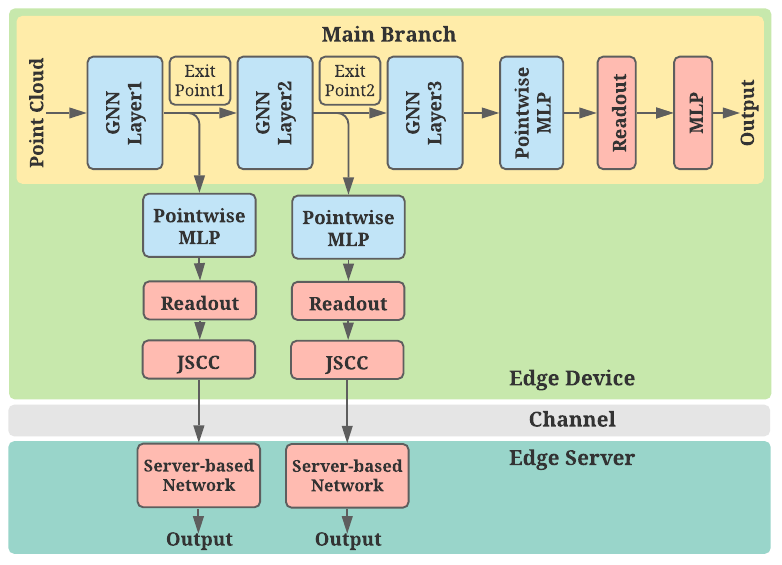}
\label{Branchy-GNN framework}
\end{minipage}%
}%
\centering
\caption{Two frameworks for device-edge co-inference: The output of each blue block is a point cloud, and the output of each red block is a fix-sized vector. 
The Split Point in (a) corresponds to the Exit Point2 in (b).}
\label{two_frameworks}
\end{figure}

\subsection{Accelerating GNN-based Edge Inference}
The common practices of edge inference are either device-only inference or edge-only inference. 
Unfortunately, both approaches may suffer from high end-to-end latency due to the constrained on-device resources and limited bandwidth.
Device-edge co-inference is a promising alternative, and \emph{model splitting} is a common method, which splits a standard network into two parts and deploys them on a device and an edge server, respectively. 
However, as investigated in the DNN-based tasks, simply splitting DNNs without altering their structures may not achieve inference acceleration in most cases because some layers in the models amplify the input data size and cause an even high communication overhand than the raw input data \cite{kang2017neurosurgeon}.
To resolve this data amplification issue, a bottleneck \cite{eshratifar2019bottlenet}, e.g., an autoencoder, is introduced to compress the intermediate feature.
Bottlenet++ \cite{shao2020bottlenet++} further adopts joint source-channel coding (JSCC) within the autoencoder to increase the coding rate, where the encoder maps the feature values to the channel symbols and incorporates the channel noise in the training process.

However, GNN-based edge inference is much more challenging than DNN counterparts.
First, GNNs suffer a more severe data amplification effect than DNNs, as the point feature would be mapped to the high-dimensional space from the 3D coordinate.
Second, building an autoencoder for large-scale graphs is difficult.
While the graph autoencoder (GAE) \cite{kipf2016variational_gae} can handle small-scale graphs (up to the order of tens), it is hard to generate the graph data of point clouds (in the order of thousands).
Even though some works \cite{yang2018foldingnet_point_cloud_1,achlioptas2018learning_point_cloud_2} attempt to directly reconstruct unordered point clouds by autoencoders, the extra computational cost is intolerable. 
Third, leveraging JSCC on graph data is more challenging, as the incorporated channel noise may influence the graph dependency between the adjacency matrix and the edge/node features, which may seriously impair the reconstruction quality.
To handle the difficulties above, we propose the Branchy-GNN framework to facilitate point cloud processing at the network edge.


\section{The Proposed Branchy-GNN Framework}
\label{proposed}

We propose Branchy-GNN as a device-edge co-inference framework for low-latency GNN inference, which modifies the original GNN by selecting the exit points and adding extra branch structures.
As shown in Fig. \ref{Branchy-GNN framework}, the exit points are selected behind the GNN layers.
In designing the Branchy-GNN framework, we address two considerations, i.e., the branch structure and the coding scheme for transmission.
In the following, we present the architecture of Branchy-GNN and illustrate its effectiveness in the point cloud classification task.

\subsection{Architecture}
The original GNN, as the main branch, introduces high computational costs from the GNN layers.
The branch network in the Branchy-GNN framework allows early exit from the main branch to reduce the on-device computation latency.
For each branch network, it consists of a pointwise MLP \cite{qi2017pointnet}, a readout layer \cite{wang2019dynamic_DGCNN}, a joint source-channel coding (JSCC) module, and a server-based network.
The first three modules are deployed on an edge device, and the last one is on an edge server.
The \textbf{pointwise MLP} applies a lightweight shared MLP for each point in the point cloud to extract the features and maintain the inference performance.
As this layer processes each point independently and does not aggregate the features from the neighbor points, it introduces far less computation complexity compared with the GNN layers in the main branch. 
Then, the \textbf{readout layer} maps the point cloud to a fixed-size vector via the sum/mean pooling, which heavily shrinks the data volume.
In particular, we adopt the readout function with the formulation $s=\frac{1}{N} \sum_{i=1}^{N} x_{i} \| \max _{i=1}^{N} x_{i}$, where $x_{i}$ is the feature vector of $i$-th point, and $N$ is the number of nodes.
Notation $\|$ denotes concatenation.
To further reduce the communication overhead, \textbf{joint source-channel coding} (JSCC) is adopted to compress the intermediate feature before transmission.
Unlike the hand-craft design of separated source and channel coding (e.g., the communication modules in Fig. \ref{model splitting}), we learn the coding scheme by a lightweight autoencoder in an end-to-end manner, where the communication channel is modeled as a non-trainable layer within the neural network. In the experiment, we consider the additive white Gaussian noise (AWGN) channel in communication. It has the transfer function $f(x) = x + n$, where $x$ is the encoded representation, and the noise $n$ is sampled from the Gaussian distribution with variance $\sigma^{2}$.
The channel condition is described by the signal-to-noise ratio (SNR), i.e., $\mathrm{SNR} = \frac{P}{\sigma^{2}} \, (\mathrm{dB})$, where $P$ is the average power of the signal.
As the encoded symbols should meet the power constraint, the output of JSCC is normalized based on the $l_{2}$ norm.
The received feature vector will be further processed at the powerful edge server.
Particularly, in classification tasks, the server-based network consists of several fully connected layers.

Compared with the model splitting framework, Branchy-GNN is less affected by data amplification.
The model splitting framework requires to explicitly restore the intermediate point cloud to maintain the inference performance, and thus it requires high-reliable communication with a large number of bits to encode the input and protect the bitstream against the channel noise.
On the contrary, the branch network in our Branchy-GNN shrinks the intermediate feature by the readout layer and the JSCC module, and the server-based network directly leverages the received data for further processing without the need for point cloud reconstruction.
Since the branch network is trained in an end-to-end manner, the server-based network can adapt to the corrupted data while maintaining the inference performance.
This is a key architecture difference compared with the model splitting framework, which deploys part of the original GNN layers on the server.




\subsection{Communication-Computation Tradeoff}

The end-to-end co-inference latency mainly consists of on-device computation latency and communication latency. The computational cost is dominated by the GNN layers in the main branch before the selected exit point, and the communication overhead is determined by the JSCC output dimension.
Note that the communication overheads in different branches are different. As the neural networks in the main branch can extract high-level information of input data and discard nuisance, it facilitates the JSCC module to encode a more compressed representation while maintaining the network performance.
As discussed above, earlier exit means less on-device computational cost and higher communication overhead than a deeper exit point.
Therefore, the exit point selection makes a communication-computation tradeoff in edge inference, and finely selecting the exit point can reduce the end-to-end latency under a specific edge environment.


\begin{figure*}[t]
\centering
\begin{minipage}[t]{0.323\linewidth}
    \centering
    \includegraphics[width=0.9\linewidth]{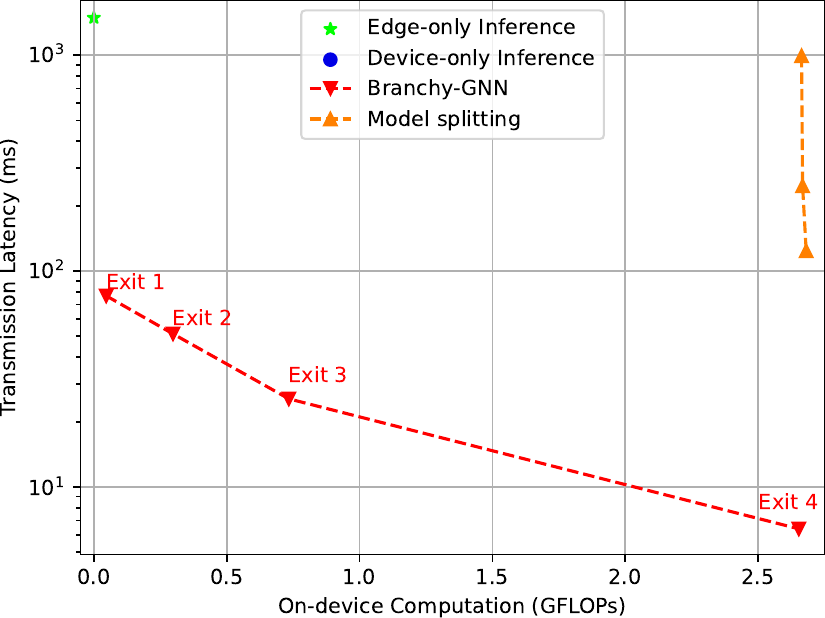}
    \caption{On-device computation cost and communication latency.}
    \label{Communication-computation Plane}
\end{minipage}%
\hspace{0.01\linewidth}
\begin{minipage}[t]{0.323\linewidth}
    \centering
    \includegraphics[width=0.9\linewidth]{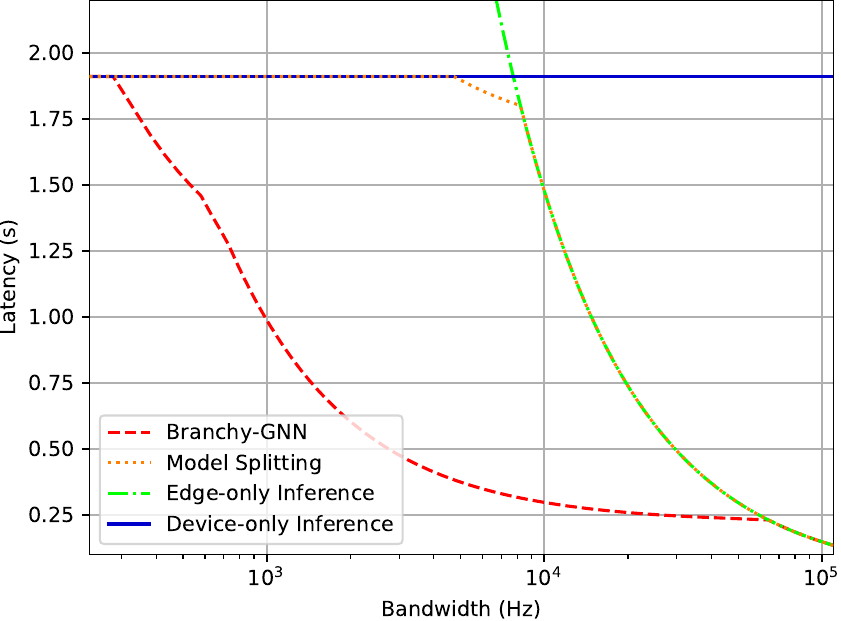}
    \caption{Edge inference latency in different bandwidth.}
    \label{bandwidth}
\end{minipage}%
\hspace{0.01\linewidth}
\begin{minipage}[t]{0.323\linewidth}
    \centering
    \includegraphics[width=0.9\linewidth]{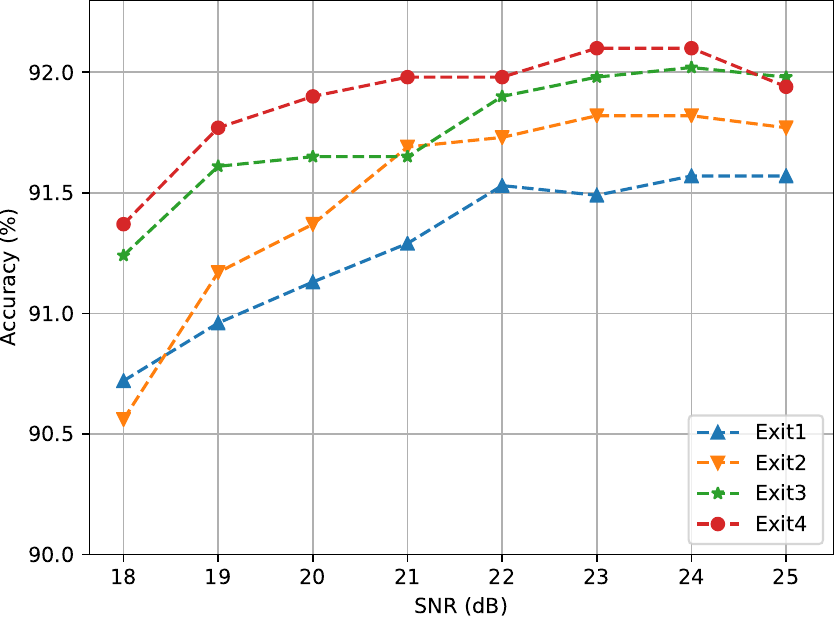}
    \caption{Classification accuracy under different channel conditions.}
    \label{robust}
\end{minipage}
\centering
\label{ablation}
\end{figure*}

\subsection{Training Methodologies}
Training the complete Branchy-GNN with multiple branches may suffer slow convergence.
To resolve this challenge, we first train the main branch, i.e., the original GNN, to the desired inference performance.
Then, we fix the network parameters in the main branch and train the multiple branch networks.
Particularly, in classification tasks, the output of the main branch can be used as soft labels for the branch networks training.
Finally, the whole Branchy-GNN framework will be fine-tuned for further performance improvement. In this step, the sum of the loss functions (e.g., cross-entropy) of each branch is used as the training objective.

\section{Experimental Results}
\label{experiment}

\subsection{Experimental Setting}

We evaluate the proposed Branchy-GNN framework on ModelNet40 \cite{wu20153d_shapenet} for a point cloud classification task, which is to predict the categories for each point cloud sample.
The dataset contains meshed computer-aided design (CAD) models from 40 categories, and we follow the experimental setting in \cite{wang2019dynamic_DGCNN} to uniformly sample 1024 points for each mesh face, where each point cloud is parameterized by the 3D coordinates and rescaled to the unit sphere.
The proposed Branchy-GNN adopts DGCNN \cite{wang2019dynamic_DGCNN} as the main branch, which consists of 4 GNN layers.
Then, we add one branch behind each GNN layer.
We require the classification accuracy close to the main branch performance (larger than 91\%) in the experiments.
Since the dimensions of the transmitted feature impact the final inference performance, we choose the dimensions from the first to the last branch as 1536, 1024, 512, 128, respectively. 
We assume the AWGN channel is with SNR of 20 dB unless otherwise specified.

To verify the effectiveness of our proposed Branchy-GNN, we compare with three different baselines in the experiments: (1) the device-only inference; (2) the edge-only inference; (3) the model splitting.
Note that the device-only and edge-only inference can be regarded as two special cases in Branchy-GNN and model splitting. The code is available at \url{https://github.com/shaojiawei07/Branchy-GNN}

\subsection{Edge Inference Speedup}

We first investigate the end-to-end edge inference latency of different edge inference schemes.
The on-device computational cost is measured by the number of floating-point operations (FLOPs), and the transmission latency reflects the communication overhead.
Note that the proposed Branchy-GNN leverages the learning-based joint source-channel coding for transmission, while other baselines assume the channel capacity $C=W\log _{2}(1+\mathrm{SNR})$ as their data rate.

Fig. \ref{Communication-computation Plane} shows the relationship between the on-device computational cost and communication latency, where the channel bandwidth $W$ is set to be $10\mathrm{kHz}$.
It can be observed that compared with the three baseline schemes, our Branchy-GNN achieves a significant saving in both on-device computational cost and communication overhead.
Besides, the multiple exit points provide a better computation-communication tradeoff compared with the model splitting method.
We then investigate the impacts of the channel bandwidth on the edge inference latency through real-life measurements.
In particular, we use a Raspberry Pi 3 as an edge device and a PC with RTX 2080 Ti as an edge server.
Fig. \ref{bandwidth} shows the measured end-to-end inference latency.
The edge-only inference would be the optimal method with sufficient bandwidth. 
However, in the frequency range of 300Hz to 60kHz, the Branchy-GNN largely reduces the inference latency compared with other baselines.



\subsection{Robustness of the Branchy-GNN}

To verify the robustness of the proposed Branchy-GNN framework, we simulate its inference accuracy at different channel conditions. The model parameters are obtained by training Branchy-GNN at the SNR of 20 dB, and directly applied at different SNR values in edge inference.
It is shown in  Fig. \ref{robust} that the classification accuracy increases with the value of SNR since the edge server can receive a more reliably intermediate feature under a better channel condition.
Besides, all the branches in Branchy-GNN maintain the classification accuracy close to or higher than 91\%, which implies that it is robust against the variation of wireless channels.

\section{CONCLUSIONS}
\label{conclusion}

In this paper, we propose Branchy-GNN for efficient point cloud processing.
It leverages branch networks and joint source-channel coding to reduce the on-device computation cost and intermediate feature transmission overhead, respectively.
Experimental results show that the proposed framework can secure much lower inference latency than other baselines.

\bibliographystyle{./IEEEbib}
\bibliography{strings,refs}

\begin{thebibliography}{10}

\bibitem{krizhevsky2012imagenet}
A.~Krizhevsky, I.~Sutskever, and G.E. Hinton,
\newblock ``Imagenet classification with deep convolutional neural networks,''
\newblock in {\em Advances in Neural Information Processing Systems}, 2012, pp.
  1097--1105.

\bibitem{maturana2015voxnet}
D.~Maturana and S.~Scherer,
\newblock ``Voxnet: A 3d convolutional neural network for real-time object
  recognition,''
\newblock in {\em 2015 IEEE International Conference on Intelligent Robots and
  Systems}, 2015, pp. 922--928.

\bibitem{riegler2017octnet}
G.~Riegler, Ali Osman~Ulusoy, and A.~Geiger,
\newblock ``Octnet: Learning deep 3d representations at high resolutions,''
\newblock in {\em Proceedings of the IEEE Conference on Computer Vision and
  Pattern Recognition}, 2017, pp. 3577--3586.

\bibitem{qi2017pointnet}
C.R. Qi, H.~Su, K.~Mo, and L.J. Guibas,
\newblock ``Pointnet: Deep learning on point sets for 3d classification and
  segmentation,''
\newblock in {\em Proceedings of the IEEE Conference on Computer Vision and
  Pattern Recognition}, 2017, pp. 652--660.

\bibitem{qi2017pointnet++}
C.R. Qi, L.~Yi, H.~Su, and L.J. Guibas,
\newblock ``Pointnet++: Deep hierarchical feature learning on point sets in a
  metric space,''
\newblock in {\em Advances in Neural Information Processing Systems}, 2017, pp.
  5099--5108.

\bibitem{wang2019dynamic_DGCNN}
Y.~Wang, Y.~Sun, Z.~Liu, S.E. Sarma, M.M. Bronstein, and J.M. Solomon,
\newblock ``Dynamic graph cnn for learning on point clouds,''
\newblock {\em Acm Transactions On Graphics}, vol. 38, no. 5, pp. 1--12, 2019.

\bibitem{diaz2016indoor}
L.~D{\'\i}az-Vilari{\~n}o, P.~Boguslawski, K.~Khoshelham, H.~Lorenzo, and
  L.~Mahdjoubi,
\newblock ``Indoor navigation from point clouds: 3d modelling and obstacle
  detection,''
\newblock {\em The International Archives of Photogrammetry, Remote Sensing and
  Spatial Information Sciences}, vol. 41, pp. 275, 2016.

\bibitem{yue2018lidar_driving}
X.~Yue, B.~Wu, S.A. Seshia, K.~Keutzer, and A.L. Sangiovanni-Vincentelli,
\newblock ``A lidar point cloud generator: from a virtual world to autonomous
  driving,''
\newblock in {\em Proceedings of the 2018 ACM on International Conference on
  Multimedia Retrieval}, 2018, pp. 458--464.

\bibitem{chen2019learning_shape_modeling}
Z.~Chen and H.~Zhang,
\newblock ``Learning implicit fields for generative shape modeling,''
\newblock in {\em Proceedings of the IEEE Conference on Computer Vision and
  Pattern Recognition}, 2019, pp. 5939--5948.

\bibitem{shao2020communication}
J.~Shao and J.~Zhang,
\newblock ``Communication-computation trade-off in resource-constrained edge
  inference,''
\newblock {\em arXiv preprint arXiv:2006.02166}, 2020.

\bibitem{shi2020communication_jun}
Y.~Shi, K.~Yang, T.~Jiang, J.~Zhang, and K.B. Letaief,
\newblock ``Communication-efficient edge ai: Algorithms and systems,''
\newblock {\em arXiv preprint arXiv:2002.09668}, 2020.

\bibitem{li2018edge_Edgenet}
E.~Li, Z.~Zhou, and X.~Chen,
\newblock ``Edge intelligence: On-demand deep learning model co-inference with
  device-edge synergy,''
\newblock in {\em Proceedings of the 2018 Workshop on Mobile Edge
  Communications}, 2018, pp. 31--36.

\bibitem{Shi2019ImprovingDC_2step}
W.~Shi, Y.~Hou, S.~Zhou, Z.~Niu, Y.~Zhang, and L.~Geng,
\newblock ``Improving device-edge cooperative inference of deep learning via
  2-step pruning,''
\newblock {\em IEEE INFOCOM 2019 - IEEE Conference on Computer Communications
  Workshops}, pp. 1--6, 2019.

\bibitem{kang2017neurosurgeon}
Y.~Kang, J.~Hauswald, C.~Gao, A.~Rovinski, T.~Mudge, J.~Mars, and L.~Tang,
\newblock ``Neurosurgeon: Collaborative intelligence between the cloud and
  mobile edge,''
\newblock {\em ACM SIGARCH Computer Architecture News}, vol. 45, no. 1, pp.
  615--629, 2017.

\bibitem{bourtsoulatze2019deepJSCC_deniz}
E.~Bourtsoulatze, D.B. Kurka, and D.~G{\"u}nd{\"u}z,
\newblock ``Deep joint source-channel coding for wireless image transmission,''
\newblock {\em IEEE Transactions on Cognitive Communications and Networking},
  vol. 5, no. 3, pp. 567--579, 2019.

\bibitem{shi2020point_gnn}
W.~Shi and R.~Rajkumar,
\newblock ``Point-gnn: Graph neural network for 3d object detection in a point
  cloud,''
\newblock in {\em Proceedings of the IEEE Conference on Computer Vision and
  Pattern Recognition}, 2020, pp. 1711--1719.

\bibitem{eshratifar2019bottlenet}
A.E. Eshratifar, A.~Esmaili, and M.~Pedram,
\newblock ``Bottlenet: A deep learning architecture for intelligent mobile
  cloud computing services,''
\newblock in {\em 2019 IEEE International Symposium on Low Power Electronics
  and Design}, 2019, pp. 1--6.

\bibitem{shao2020bottlenet++}
J.~Shao and J.~Zhang,
\newblock ``Bottlenet++: An end-to-end approach for feature compression in
  device-edge co-inference systems,''
\newblock in {\em 2020 IEEE International Conference on Communications
  Workshops}, 2020, pp. 1--6.

\bibitem{kipf2016variational_gae}
T.N. Kipf and M.~Welling,
\newblock ``Variational graph auto-encoders,''
\newblock {\em NIPS Workshop on Bayesian Deep Learning}, 2016.

\bibitem{yang2018foldingnet_point_cloud_1}
Y.~Yang, C.~Feng, Y.~Shen, and D.~Tian,
\newblock ``Foldingnet: Point cloud auto-encoder via deep grid deformation,''
\newblock in {\em Proceedings of the IEEE Conference on Computer Vision and
  Pattern Recognition}, 2018, pp. 206--215.

\bibitem{achlioptas2018learning_point_cloud_2}
P.~Achlioptas, O.~Diamanti, I.~Mitliagkas, and L.~Guibas,
\newblock ``Learning representations and generative models for 3d point
  clouds,''
\newblock in {\em International Conference on Machine Learning}, 2018, pp.
  40--49.

\bibitem{wu20153d_shapenet}
Z.~Wu, S.~Song, A.~Khosla, F.~Yu, L.~Zhang, X.~Tang, and J.~Xiao,
\newblock ``3d shapenets: A deep representation for volumetric shapes,''
\newblock in {\em Proceedings of the IEEE Conference on Computer Vision and
  Pattern Recognition}, 2015, pp. 1912--1920.

\end{thebibliography}

\end{document}